# Calculations of gamma-ray spectral profiles of linear alkanes in the positron annihilation process

Xiaoguang Ma[1,2] and Feng Wang[*,1]

[1]*eChemistry Laboratory, Department of Chemistry and Biotechnology, School of Science, Faculty of Science, Engineering and Technology, Swinburne University of Technology, PO Box 218, Hawthorn, Victoria 3122, Australia*

[2]*School of Physics and Optoelectronic Engineering, Ludong University, Yantai, Shandong 264025, PR China*

The positron-electron annihilation gamma-ray spectra of linear alkanes $C_nH_{2n+2}$ (n=1-12) have been studied systematically. A profile quality (PQ) parameter, $\chi$, is introduced to assess the agreement between the obtained theoretical profiles and the experimental measurements in the entire region of energy shift of the spectra. Together with the Doppler shift ($\Delta\varepsilon$) of the gamma-ray spectra, the two parameters, $\chi$ and $\Delta\varepsilon$, are able to provide a more comprehensive assessment of the calculated gamma-ray spectra with respect to available experiment. Applying the recently developed docking model, the present study determines the positrophilic electrons for individual alkanes from which the gamma-ray spectral profiles are calculated. The results achieve an excellent agreement with experiment, not only with respect to the Doppler shift, but also with respect to the gamma-ray profiles in the photon energy region up to 5 keV. The study further calculates the gamma-ray spectra of other linear alkanes in the series without available experimental measurements, such as heptane ($C_7H_{16}$), octane ($C_8H_{18}$), decane ($C_{10}H_{22}$) and undecane ($C_{11}H_{24}$). The results obtained show a dominance of the positrophilic electrons in the lowest occupied valence orbital (LOVO) in the positron-electron annihilation process, in agreement with previous studies.

# 1. INTRODUCTION

Recently, it was theoretically revealed that specific valence electrons, i.e., the positrophilic electrons in methane and hexane, dominate their annihilation spectra in gas phase.[1,2] Under the low-energy positron plane wave positron (LEPWP) approximation,[7] the $2a_1$ electrons of methane[1] are identified as the positrophilic electrons which dominantly locate in the lowest occupied valence orbital (LOVO).[2] A further study of polar molecules, such as fluorinated methanes,[3] showed that the LOVO electrons again dominate the contribution of the positrophilic electrons in the molecules, in agreement with the results obtained for non-polar molecules. As a result, the docking model which determines the positrophilic electrons from valence electrons of a molecule was introduced.[3] The calculated gamma-ray spectra of the molecules, polar and non-polar, using the docking model agreed well with the experiments.[3] It was further indicated that these positrophilic electrons which are dominated by the LOVO electrons of a molecule play an important role in positron-electron annihilation process.

In recent studies,[1-3] it is demonstrated that under the LEPWP approximation,[7] the positrophilic electrons are determined by docking the electron density distributions into the local molecular attraction potential (LMAP) of the molecule, both of the electron densities and LMAP are calculated quantum mechanically.[3] In this docking model, the influence of positron wave function is very small and therefore, can be neglected and the LMAP becomes the dominant term to reflect the molecular force field effect on the positron wavefunction.[3] The electrons docking into the positron attractive potential possess the dominant probabilities to annihilate a positron.

The present study applies the docking model under the LEPWP approximation to determine positrophilic electrons, in order to calculate the $\gamma-$ray positron-electron annihilation spectra of a completed series of linear alkane molecules in gas phase, i.e., $C_nH_{2n+2}$ (n=1-12), for which accurate measurements are available for many of the molecules.[4] Theoretical expressions of the gamma-ray spectra in the positron-electron annihilation process are given in the section 2. The calculated results of the linear alkanes are compared with available measurements to validate the theoretical model. The model has been applied to study alkanes in the series which do not have experimental measurement available in section 3, and finally, the conclusions are drawn in section 4.

## 2. THEORETICAL TREATMENT

The local molecular attraction potential (LMAP) which is presented to accommodate the docking in the positron-electron annihilation process, is given by[3]

$$U(r) = -\rho_p(r) \times V_{mol}(r). \quad (1)$$

Where $\rho_p(r)$ is the positron density and $V_{mol}(r)$ is the total electrostatic potential, which are calculated quantum mechanically. The local molecular attraction potential (LMAP) is used to describe the physical interaction between a positron and a molecule.

The total electrostatic potential of a molecule represents electrostatic Coulomb interactions with a positive unit charge in a molecule, which includes nuclear $V_{nuc}(r)$ and electronic $V_{ele}(r)$ interactions in a molecule,

$$V_{mol}(r) = V_{nuc}(r) + V_{ele}(r) = \sum_A \frac{Z_A}{|r-R_A|} - \int \frac{\rho_e(r')}{|r-r'|} dr'. \quad (2)$$

Substitution of $V_{mol}(r)$ into Eq.(1), the LMAP $U(r)$ becomes

$$U(r) = -\rho_p(r) \cdot V_{nuc}(r) - \rho_p(r) \cdot V_{ele}(r)$$
$$= -\sum_A \frac{\rho_p(r) \cdot Z_A}{|r-R_A|} + \int \frac{\rho_p(r) \cdot \rho_e(r')}{|r-r'|} dr', \quad (3)$$

where the $\rho_e(r')$ is the electron density. Under the LEPWP approximation,[7] the positron density $\rho_p(r) \approx 1$. As a result, the LMAP only depends on $V_{mol}(r)$ for the positrophilic and electrophilic sites on the molecule in the annihilation processes[3]

$$U(r) = -\sum_A \frac{Z_A}{|r - R_A|} + \int \frac{\rho_e(r')}{|r - r'|} dr'. \tag{4}$$

Under the Born-Oppenheimer approximation, the wave function of the positrophilic electrons in the $i$th orbital can be expanded by Gaussian type functions (GTFs)

$$\psi_i(r) = \sum_{klmn} C_{klmn} x^k y^l z^m \exp(-a_n r^2). \tag{5}$$

where the expansion coefficient $C_{klmn}$ represents the $i$th molecular orbital obtained by self-consistent methods and $x^k y^l z^m \exp(-a_n r^2)$ are basis functions.

Therefore, the electron density can be calculated as,

$$\rho_e(r) = \sum_i \eta_i |\psi_i(r)|^2, \tag{6}$$

where $\psi_i(\mathbf{r})$ is the wave function of the electron in the $i$th orbital of the target in the ground electronic state, and the positron density is given by,

$$\rho_p(r) = \sum_k \eta_k |\varphi_k(r)|^2. \tag{7}$$

here $\varphi_\mathbf{k}(\mathbf{r})$ is the wave function of the incident positron with momentum $\mathbf{k}$ and $\eta$ is the occupation number.

The real space wave functions are then directly mapped into the momentum space[6]

$$A_{i\mathbf{k}}(\mathbf{P}) = \int \psi_i(\mathbf{r}) \varphi_\mathbf{k}(\mathbf{r}) e^{-i\mathbf{P}\cdot\mathbf{r}} d\mathbf{r}. \tag{8}$$

Where **P** is the total momentum of the annihilation photons. The probability distribution function at the photon momentum **P** in two-photon annihilation is then given by

$$W_i(\mathbf{P}) = \pi r_0^2 c |A_{i\mathbf{k}}(\mathbf{P})|^2, \qquad (9)$$

where $r_0$ is the classical electron radius, $c$ is the speed of light. The spherically averaged gamma-ray spectra for each type of electrons are then calculated by using general equations,[6,7]

$$w_i(\varepsilon) = \frac{1}{c} \int \int_{2|\varepsilon|/c}^{\infty} W_i(\mathbf{P}) \frac{PdPd\Omega_\mathbf{P}}{(2\pi)^3}. \qquad (10)$$

The Doppler shift is therefore, calculated using Eq.(10).

Gamma-ray spectrum of a molecule is a profile over a photon energy range of up to $\Delta\varepsilon$ (5 keV) shift from the central 511 keV position (**P**=0) when P≠0. Certainly, the Doppler shift ($\Delta\varepsilon$, i.e, the full width at half maximum (FWHM)) has been recognised as the most important parameter to quantify the spectra of molecules[1-7] and is closely correlated to electron momentum distribution of the molecule.[7] However, the full width at half maximum (FWHM, $\Delta\varepsilon$) parameter can be a necessary but not a sufficient condition to ensure the goodness of the calculated momentum profiles. For a satisfied agreement between theory and measurement, in this study, we introduced the second parameter, i.e., profile quality ($\chi$) which is given by the root-mean square deviation (RMSD) between the theoretical and the experimental results of the gamma-ray spectrum of a molecule,

$$\chi = \sqrt{\frac{\sum_{i=1}^{N}(\sigma_{T,i} - \sigma_{E,i})^2}{N}} \times 100\%. \qquad (11)$$

Where $\sigma_{E,i}$ is the points in the two-Gaussian (2G) fitted experimental gamma-ray spectrum, while $\sigma_{T,i}$ is the corresponding theoreticaly calculated profile. Here N is the total number of the data points, which is given by N=3001 in the present study for up to 5 keV. This number of points is a sufficiently large number to represent the agreement of the gamma-ray spectral profile. All spectra are normalized to one in arbitrary unit at 511 keV and therefore, all energy shifts are relative to the position at 511 keV.

Theoretical calculations for the electronic structures of alkanes are *ab initio* Hartree-Fock (HF) calculations, which are combined with the TZVP basis set.[8] In this basis set atomic carbon orbitals are constructed by the *5s9p6d* scheme of Gaussian type functions (GTFs), while atomic hydrogen orbitals are constructed by the *3s3p* scheme of GTFs. The molecular wave functions are then obtained quantum mechanically using the HF/TZVP model and based on the optimised structural parameters.[9,10, 13] All electronic structural calculations are implemented using the Gaussian09 computational chemistry package.[11]

## 3. RESUTLS AND DISCUSSIONS

In the series of linear alkanes, $C_nH_{2n+2}$ (n=1-12), accurate gamma-ray experimental measurements[4] are available for methane to hexane (n=1-6), nonane (n=9) and dodecane (n=12), whereas the alkanes with n=7, 8, 10 and 11, that is, heptane ($C_7H_{16}$), octane ($C_8H_{18}$), nonane ($C_9H_{20}$) and undecane ($C_{11}H_{24}$) do not have such the measurements. As a result, the theoretical model will be validated using the available experimental results before being applied to study the alkanes without measurements.

Our previous studies[1-3,7,10] have clearly indicated that the contributions to the Doppler shfit of gamma-ray in a molecule is orbital dependent. For example, the $\gamma$–ray spectra of the smallest alkane molecule, CH$_4$,[1] reveal that the positrophilic electrons are dominated by the 2a$_1$ electrons in the LOVO of this molecule. The Doppler shift FWHM ($\Delta\varepsilon$) of the gamma-ray profile of the 2a$_1$ electrons of methane is given by 2.02 keV, while the experimental result is 2.06 keV. The difference of $\Delta\varepsilon$ is 0.04 keV. However, the overall profile quality, $\chi$, of the gamma-ray profile between the calculation and the measurement is given by 0.27% (<<1), indicating that an excellent overall agreement in the energy region is achieved. Therefore, the electrons in the LOVO of methane dominate the electron-positron annihilation process under the LEPWP approximations.[1,7] In a same vein, the gamma-ray spectra of hexane are studied recently.[2] It was found in gas phase that the positrophilic electrons of hexane are from a number of valence orbitals but all are under the highest occupied molecular orbital (HOMO), in agreement with an earlier study of Crawford.[12]

Figure 1 compares the $\gamma$– ray spectral profiles of other alkanes which have available experimental results,[4] that is, ethane (C$_2$H$_6$), propane (C$_3$H$_8$), butane (C$_4$H$_{10}$), pentane (C$_5$H$_{12}$), nonane (C$_9$H$_{20}$) and dodecane (C$_{12}$H$_{26}$), where the optimised molecular structures of the alkanes are also given in the figure. The experimental results [4] of two-Gaussian (2G) fit are employed. Two spectral profiles are calculated, one for all positrophilic electrons and the other only for the electrons in the LOVO orbital of the alkanes. The first number in brackets is the full widths at half maximum (FWHM), i.e., the Doppler shift ($\Delta\varepsilon$) in keV unit and the second number is the profile quality ($\chi$) in percentage. The positrophilic electrons of the alkanes are using the docking mechanism.[3] All spectra in this figure are normalized to unity at zero (511 keV).

Table 1 compares the calculated spectral profiles with the available measurements. One set of calculations includes all positrophilic electrons determined using the docking model,[3] the other set of calculations include the LOVO electrons only. The calculated spectra of theses alkanes are in excellent agreement with available experimental measurements. For example, the profile quality, $\chi$, indicates that the largest discrepancy between the measurement and calculation is under 0.45% in the positrophilic electron model, whereas this discrepancy becomes under 3.25% in the LOVO electrons only model. In ethane ($C_2H_6$), the docking model[2,3] indicates that the positrophilic electrons consist of valence electrons in the $2a_g$ and $2b_u$ orbitals, in which the $2a_g$ electrons are LOVO electrons. The electron densities of these positrophilic electrons are docked into the positive LMAP region. Although there are four positrophilic electrons (in two valence orbitals) in ethane, the contributions to the annihilation process are not the same. For example, if only the two LOVO ($2a_g$) electrons of ethane are considered, the Doppler shift ($\Delta\varepsilon$) of ethane is given by 1.79 keV, which results in the profile quality ($\chi$) of 1.20% in the entire photon energy region of up to 5 keV. However, if only the two positrophilic electrons in the $2b_u$ orbitals are considered, the calculated Doppler shift ($\Delta\varepsilon$) of ethane is given by 2.83 keV, which results in the profile quality ($\chi$) of 2.04% in the entire photon energy region of up to 5 keV. That is, the profile quality of the $2b_u$ electrons significantly decreases when comparing to the LOVO electrons. The latter dominates the positron-electron annihilation process of the $C_2H_6$ molecule.

The calculations in Table 1 also reveal that the number of positrophilic electrons increases as the size of the alkanes increases, although the measured Doppler shift of the alkanes does not vary significantly. For example, in methane, the positrophilic electrons are the determined by a pair of LOVO electrons ($2a_1$)[1] whereas in hexane, the positrophilic electrons locate on three different orbitals as $4a_g$ (LOVO), $6a_g$ and $6b_u$.[2] There are eight positrophilic electrons in dodecane, including the electrons in the $7a_g$ (LOVO), $9a_g$, $11b_u$ and $12a_g$ orbitals. As the size of alkane grows, the valence electrons are more delocalised so that the LOVO electrons play a less important role, although the LOVO electrons are still dominant the gamma-ray spectra of the alkanes under study.

Interestingly, the positrophilic electrons of the alkanes concentrate in the inner valence region. In Figure 1, all calculated $\gamma$–ray spectra of these alkane using the positrophilic electrons (the solid lines) agree well with the 2G fitted spectral profiles of the measurements.[4] The calculated spectral profiles of the LOVO electrons (the red dashed lines) of the alkanes under estimate the gamma-ray spectra, in general. The spectra in this figure further exhibit that the larger the size of the alkane, the larger the discrepancies between the measurements and the LOVO electrons. For example, the LOVO electrons of methane ($CH_4$) produce a Doppler shift of 1.79 keV in the $\gamma$–ray spectrum. The discrepancy ($\Delta\Delta\varepsilon$) between the $\Delta\varepsilon$ of the LOVO electrons and the measurement of methane ($C_nH_{2n+2}$, n=1) is −0.04 keV (see Table 1). This discrepancy increases to −0.37 keV, −0.53 keV, −0.74 keV, −0.75 keV, −0.76 keV (hexane, see Table 1), −0.94 keV and −1.04 keV, as the number of carbon atoms increases to n=2, 3, 4, 5, (6), 9 and 12, respectively in the linear alkanes.

Positrons are anti-matter of electrons. The process of positron-electron annihilation of molecules results in ionized molecular cations, $M^+$ and is therefore, an ionization process. However, the annihilation process is significantly different from conventional ionizations where an electron(s) leaves the molecule by, such as absorption of a photon or electron collisions, although positron-electron annihilation exhibits similarities to ionization. The calculated gamma-ray spectral profiles from the positrophilic electrons of the alkanes in Figure 1 almost reproduce the 2G fitted measurement spectral profiles,[4] as well as in excellent agreement with the Doppler shift of the measured alkanes, indicating that the present positrophilic electron model[1-3] is indeed a good model for the gamma-ray spectra of the electron-positron annihilation process of molecules. Further inspections into the positrophilic electrons in Table 1, it is revealed that not only the LOVO electrons are in the inner most valence orbital, but also all the positrophilic electrons locate in the inner valence orbitals of the alkanes, rather than in the conventional outer valence orbitals or HOMO (highest occupied molecular orbitals) as found in electron chemistry.[14] For example, the positrophilic electrons of dodecane ($C_{12}H_{26}$) locate in four inner valence orbitals of $7a_g$, $9a_g$, $11b_u$ and $12a_g$. The findings in the present study also agree with an earlier study of Crawford that "Calculations for hydrocarbons, presented below, suggest that most of the time the hole is created in an MO below the highest occupied molecular orbital (HOMO)."[12] As a result, molecules are very different from atoms where the electrons in the HOMO exhibit the largest probability to annihilate a positron.[6]

Figure 2 reports the calculated gamma-ray spectra of the alkanes in the series without measurements, that is, heptane ($C_7H_{16}$), octane ($C_8H_{18}$), decane ($C_{10}H_{22}$) and undecane ($C_{11}H_{24}$), using the same docking model. As shown in this figure, the strongest attractive potential LMAPs of heptane and octane for positron outline the carbon planes as shown in the inset contour maps. The positrophilic electrons of an alkane are determined when docking the calculated electron densities of the valence orbitals into the strong attractive LAMP of the molecule. The color-filled contour map of the positrophilic electrons (on the plane of the carbon frame) are therefore, determined from the docking. Table 2 summaries the information of these alkanes which the gamma-ray spectra are calculated.

The calculated Doppler shift of heptane ($C_7H_{16}$), octane ($C_8H_{18}$), decane ($C_{10}H_{22}$) and undecane ($C_{11}H_{24}$) are given by 2.39 keV, 2.36 keV, 2.31 keV and 2.38 keV, respectively, which is in consistent with the Doppler shift of other alkanes in this series with measurements. It is suggested that the calculated Doppler shift results can be a good predition. The positrophilic electrons are contributions from electrons in a number of valence orbitals which includes the LOVO electrons. For example, the LOVO electrons of heptane ($C_7H_{16}$) on the 8a' orbital, again dominate the contributions of positrophilic electrons. Other positrophilic electrons of heptane on the inner valence orbitals of 13a' and 14a' make minor contributions, which ensure that the spectral profiles of the alkane are in the like-hood of the reality. The calculated Doppler shift of the $\gamma$−ray spectra in the positron-electron annihilation process of heptane ($C_7H_{16}$) is given by 2.39 keV, which is in the vicinity of the calculated $\Delta\varepsilon$ of 2.44 keV for hexane ($C_6H_{14}$) and of 2.39 keV for nonane ($C_9H_{20}$). Another example is undecane ($C_{11}H_{24}$). The positrophilic electrons of undecane are electrons from four different inner valence orbitals, such as, 12a'(LOVO), 16a', 20a' and 22a'. The calculated $\Delta\varepsilon$ of the $\gamma$−ray spectrum of undecane is 2.38 keV. It is again in the vicinity of the measured $\Delta\varepsilon$ of nonane ($C_9H_{20}$) of 2.31 keV and of dodecane ($C_{12}H_{26}$) of 2.31 keV. Further experimental measurements of these alkanes are warranted to confirm the calculations in this study.

Figure 3 compares the calculated (red solid circles) with the measured (i.e., single Gaussian fitted (1G: blue starts) and two Gaussian fitted (2G: black solid squares)) Doppler shift ($\Delta\varepsilon$) of the alkanes. It is noted that small errors introduced by the analysis schemes of the measurements are observed. For example, small discrepancies are observed between the 1G and 2G fits of the same measurements. This is particularly obvious in smaller alkanes such as methane (1G: 2.09 keV and 2G: 2.06 keV), ethane (1G: 2.18 keV and 2G: 2.16 keV), propane (1G: 2.21 keV and 2G: 2.24 keV) and butane (1G: 2.28 keV and 2G: 2.31 keV).[4] Interestingly, the agreement between the 1G and 2G fits of the measurements for larger alkanes, such as hexane ($C_6H_{14}$, 2.25/2.25 keV) and nonane ($C_9H_{20}$, 2.32/2.31 keV) are very small. It is also noted that smaller discrepancy between 1G and 2G fitted measurements of dodecane ($C_{12}H_{26}$) exists as 2.29 keV and 2.31 keV, respectively,[4] whereas the agreement between the 2G fit of the measurement (2.31 keV) and the calculated (2.31 keV) gamma-ray is even better than the 1G (2.29 keV) and 2G fit of the same measurement. This figure shows that the calculated and the measured Doppler shift achieve excellent agreement in the linear alkane series. The calculated Doppler shift of the alkanes without available measurements exhibits a consistent trend in the series of linear alkanes.

There does not exist a defined relationship between the Doppler shift and the size of the alkanes (i.e., the number of carbon atoms, n, in the chains), nor the total number of the bound electrons but positrophilic electrons. For example, the measured FWHM exhibits a small increase as the increase of n for n<=4, that is, $\Delta\varepsilon$ is given by 2.06 keV, 2.16 keV, 2.24 keV and 2.34 keV, accordingly, for n=1, 2, 3 and 4 (methane, ethane, propane and butane). The trend starts to drop at n-pentane ($\Delta\varepsilon$ = 2.24 keV) and remains almost the same at n-hexane ($\Delta\varepsilon$ = 2.25 keV), but becomes stable in larger alkane such as n-noane ($\Delta\varepsilon$ = 2.31 keV) and n-dodecane ($\Delta\varepsilon$ = 2.31 keV).[4] This is similar to the HOMO-LUMO energy gaps of small alkanes where the energy gaps do not directly dependent on the size of the alkane.[13] As indicated that the Doppler shift of a bound system such as atoms and molecules, does not directly depend on the total number of bound electrons.

It is also noted that the discrepancies of Doppler shift between the measured and the calculated are apparently larger for butane, pentane and hexane (n=4-6), which indicates a that the positron wavefunction may play a more important role for theses alkanes so that inclusion of the positron wavefunction in the model is hoped to improve the agreement. As a result, it is possible that such the gamma-ray technique will have an excellent potential to study small molecules and their structures such as isomerization such as hexane and isomers [15] rather than being employed to study large alkanes (i.e., n>12).

## 4. CONCLUSIONS

The positron-electron annihilation gamma-ray spectra of linear alkanes $C_nH_{2n+2}$ (n=1-12) have been studied quantum mechanically using low energy plane wave positron approximation. [7] In order to assess the quality of the gamma-ray spectral profile in the entire photon energy region, in addition to the Doppler shift ($\Delta\varepsilon$), a profile quality (PQ) parameter, $\chi$, is introduced to measure the agreement between the obtained theoretical profiles and the experimental measurements. Applying the recently developed docking model, the present study determines the positrophilic electrons for individual alkanes from which the gamma-ray spectral profiles are calculated. The results achieve an excellent agreement with available experiment, not only with respect to the Doppler shift, but also with respect to the gamma-ray profiles in the photon energy region up to 5 keV. The present study further calculates the gamma-ray spectra of the missing linear alkanes in the series which are not measured experimentally, such as heptane (2.39 keV, $C_7H_{16}$), octane (2.36 keV, $C_8H_{18}$), decane (2.31 keV, $C_{10}H_{22}$) and undecane (2.38 keV, $C_{11}H_{24}$). The results obtained show a dominance of the positrophilic electrons in the lowest occupied valence orbital (LOVO) in the positron-electron annihilation process, in agreement with previous studies.

Additional proof: After we submitted this manuscript in September 2013, the discussion of application of the nuclear orbital plus molecular orbital (NOMO) theory [16] with Nakai group has made progress. The most recent calculations on noble gas atoms indicated that total electron-positron wave function of a bound system is dominated by the electronic wave function of the system.[17]

## ACKNOWLEDGEMENTS

This project is supported by the Australia Research Council (ARC) under the discovery project (DP) scheme. National Computational Infrastructure (NCI) at the Australia National University (ANU) under the Merit Allocation Scheme (MAS) and Swinburne University's GPU supercomputing facilities are acknowledged. Dr Ma acknowledges the support of the Natural Science Foundation Project of Shandong Province (No.ZR2011AM010).

## REFENERNCES

TABLE 1. Profile quality ($\chi\%$) and Doppler shift ($\Delta\varepsilon$) of the gamma-ray spectra of the alkanes with available experimental results.

| Alkanes | Theoretical calculated gamma-ray spectral parameters | | | | | | Expt.[*] |
| --- | --- | --- | --- | --- | --- | --- | --- |
| | LOVO electrons | | | Positrophilic electrons | | | |
| | Orbital(s) | $\chi$(%) | $\Delta\varepsilon$(KeV) | Orbital(s) | $\chi$(%) | $\Delta\varepsilon$(KeV) | $\Delta\varepsilon$(KeV) |
| Methane(CH$_4$) | 2a$_1$ | 0.27 | 2.02 | 2a$_1$ | 0.27 | 2.02 | 2.06 |
| Ethane (C$_2$H$_6$) | 2a$_g$ | 1.20 | 1.79 | 2a$_g$,2b$_u$ | 0.21 | 2.24 | 2.16 |
| Propane (C$_3$H$_8$) | 3a$_1$ | 1.74 | 1.71 | 3a$_1$,4a$_1$ | 0.34 | 2.31 | 2.24 |
| Butane (C$_4$H$_{10}$) | 3a$_g$ | 2.30 | 1.57 | 3a$_g$,4b$_u$ | 0.28 | 2.16 | 2.31 |
| Pentane (C$_5$H$_{12}$) | 4a$_1$ | 2.41 | 1.49 | 4a$_1$,6a$_1$ | 0.43 | 2.09 | 2.24 |
| Hexane (C$_6$H$_{14}$) | 4a$_g$ | 2.61 | 1.49 | 4a$_g$,6a$_g$,6b$_u$ | 0.32 | 2.44 | 2.25 |
| Nonane (C$_9$H$_{20}$) | 10a' | 3.13 | 1.34 | 10a',13a',17a',18a' | 0.27 | 2.39 | 2.31 |
| Dodecane (C$_{12}$H$_{26}$) | 7a$_g$ | 3.24 | 1.27 | 7a$_g$,9a$_g$,11b$_u$,12a$_g$ | 0.26 | 2.31 | 2.31 |

[*] see Ref.4

TABLE 2. Calculated Doppler shift of gamma-ray spectra of linear alkanes without available experimental results.

| Alkanes | Calculated Doppler shift | | | |
|---|---|---|---|---|
| | LOVO electrons | | Positrophilic electrons | |
| | Orbital(s) | $\Delta\varepsilon$(KeV) | Orbital | $\Delta\varepsilon$(KeV) |
| Heptane ($C_7H_{16}$) | 8a' | 1.42 | 8a',13a',14a' | 2.39 |
| Octane ($C_8H_{18}$) | $5a_g$ | 1.34 | $5a_g,8a_g,8b_u$ | 2.36 |
| Decane ($C_{10}H_{22}$) | $6a_g$ | 1.34 | $6a_g,7b_u,10a_g,10b_u$ | 2.31 |
| Undecane ($C_{11}H_{24}$) | 12a' | 1.27 | 12a',16a',20a',22a' | 2.38 |

Figure captions

Figure 1: Comparison of the calculated gamma-ray spectral profiles (solid line) of ethane, propane, butane, pentane, hexane, nonane and dodecane in positron-electron annihilation process with two-Gaussian fitted experimental measurements (the circles). The LOVO electron contributions are given in red dashed line.

Figure 2: The calculated gamma-ray spectra of heptane, octane, decane and undecane in positron-electron annihilation process. The solid line is given by positrophilic electrons whereas the red dashed line represents for LOVO electrons. In the upper panels of heptane and octane spectra, the local molecular attraction potential (LMAP), which docks with the electron density (DEN) where the positrophilic electrons are determined, is given.

Figure 3: Comparision of the calculated (red solid circles) Doppler shift of gamma-ray spectra of the alkanes with one-Gaussian (blue stars) and two-Gaussian (black squares) fitted experiments (KeV). The agreement with the experiment is within the experimental errors bars.

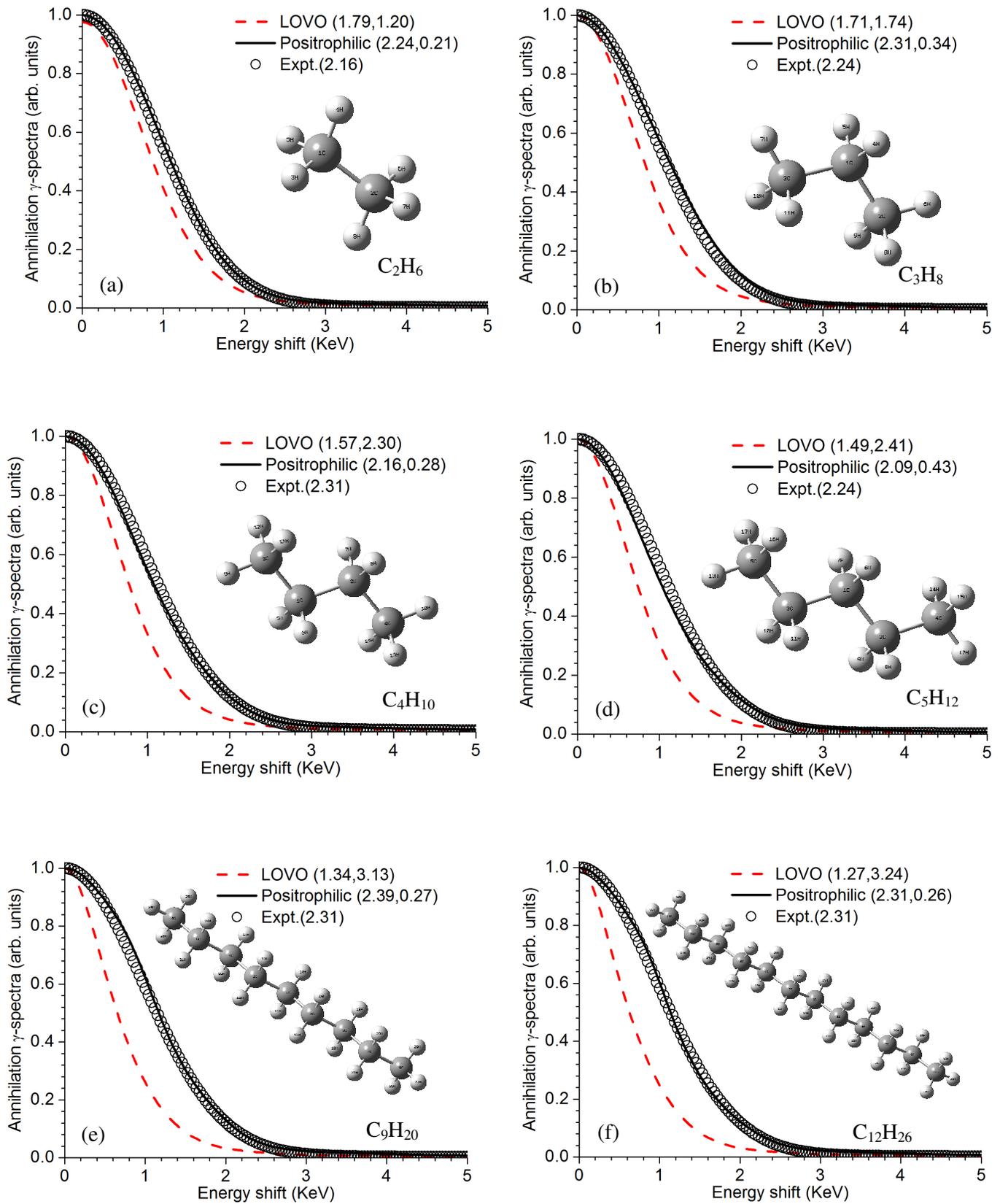

Fig.1

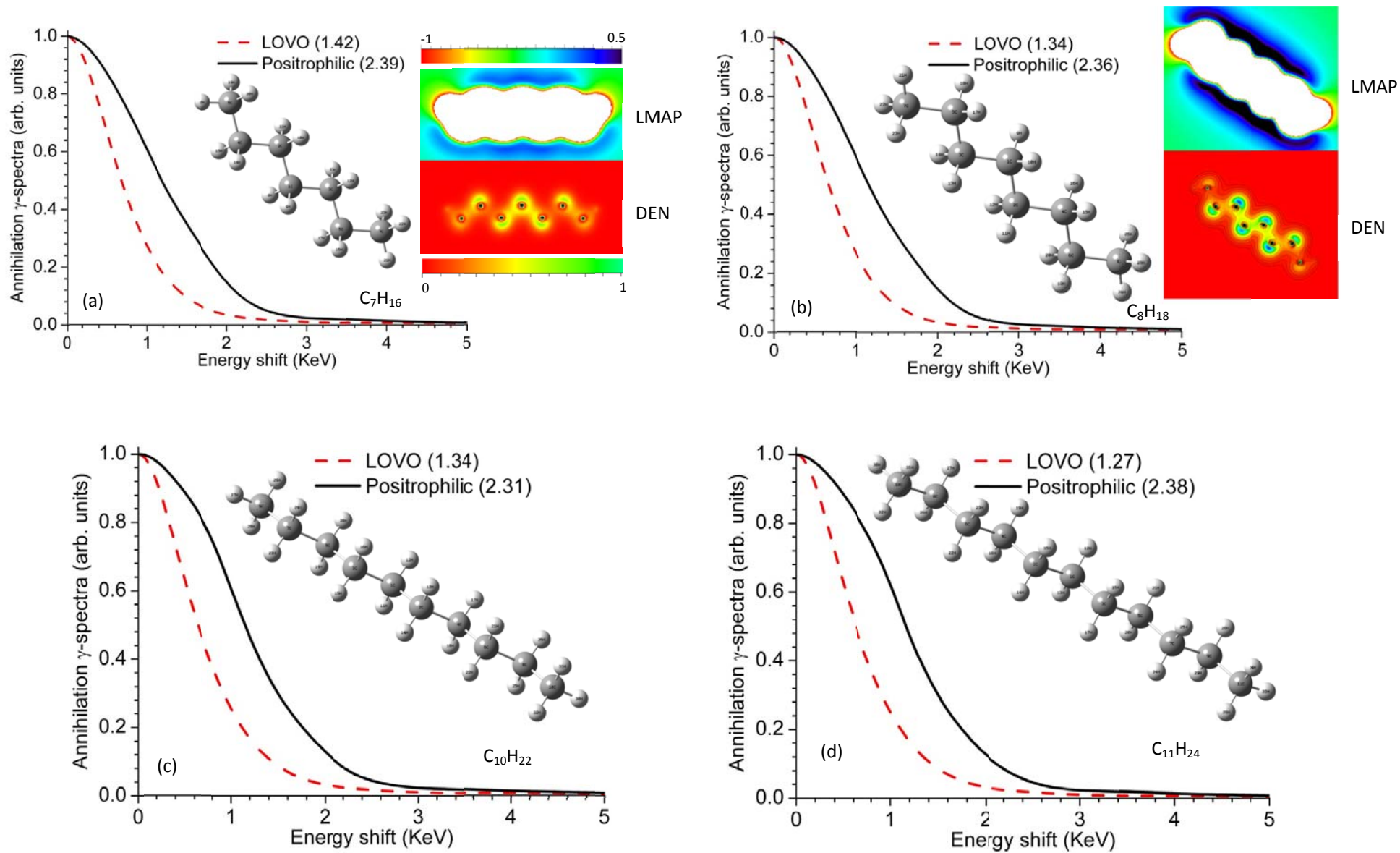

Fig.2

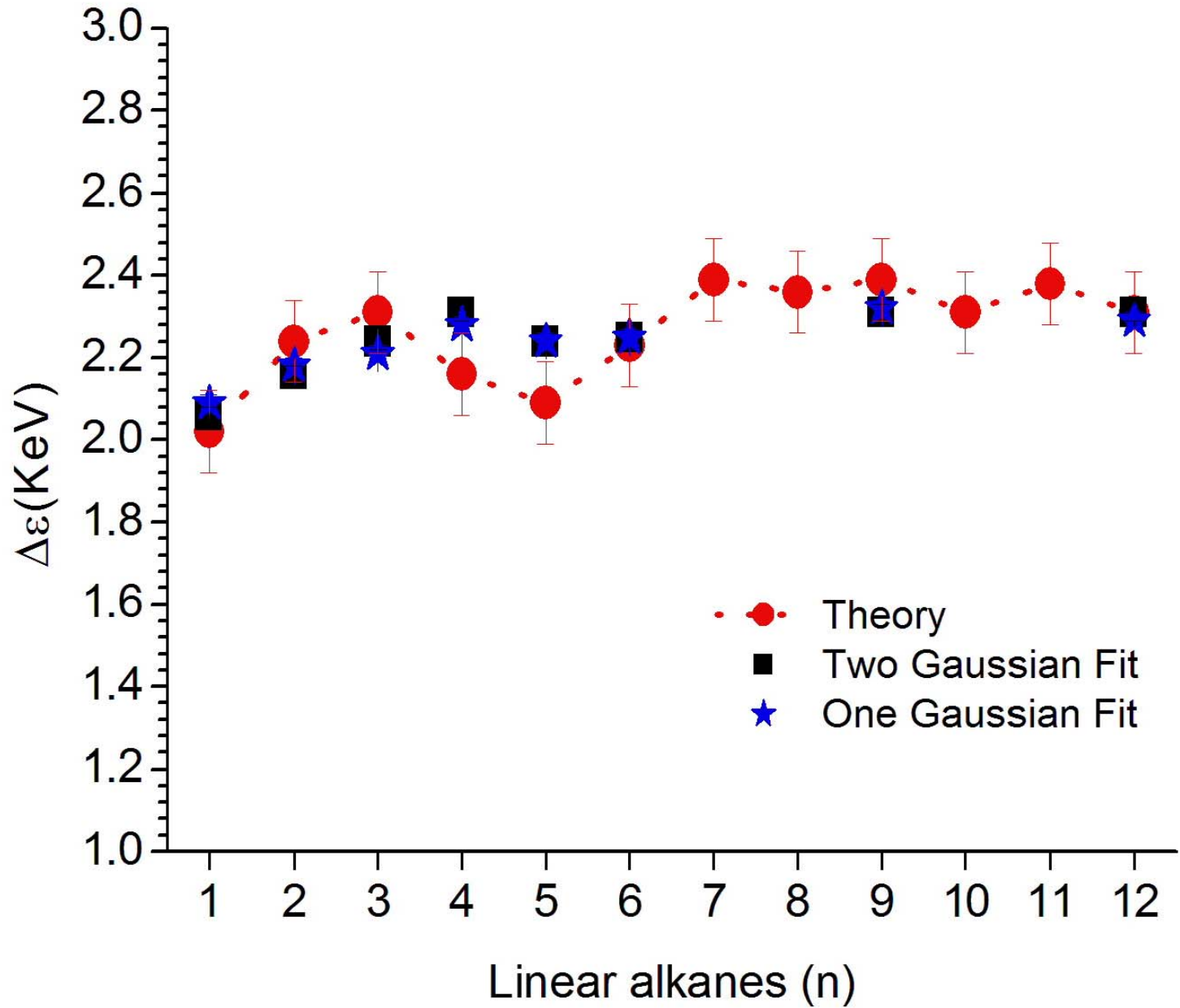

Fig.3